%% file: main.tex
\documentclass[conference]{IEEEtran}
\IEEEoverridecommandlockouts
\input{preamble}
\usepackage{cite}
\usepackage{amsmath,amssymb,amsfonts}
\usepackage{algorithmic}
\usepackage{graphicx}
\usepackage{textcomp}
\usepackage{xcolor}
\def\BibTeX{{\rm B\kern-.05em{\sc i\kern-.025em b}\kern-.08em
    T\kern-.1667em\lower.7ex\hbox{E}\kern-.125emX}}

\usepackage{listings}
\usepackage{caption}
\usepackage{xcolor}
\usepackage{float}
\usepackage{subfig}

\begin{document}

\newcommand{\orcid}[1]{\href{https://orcid.org/#1}{\includegraphics[height=10pt]{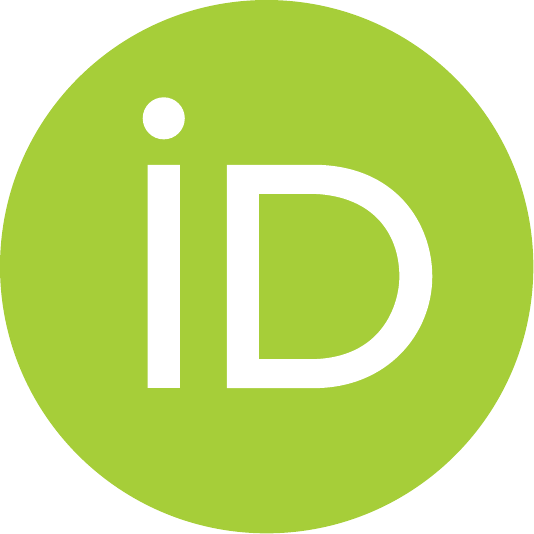}}}

\def\BibTeX{{\rm B\kern-.05em{\sc i\kern-.025em b}\kern-.08em
    T\kern-.1667em\lower.7ex\hbox{E}\kern-.125emX}}
\title{Overcoming Latency-bound Limitations of Distributed Graph Algorithms using the HPX Runtime System }

\author{
\begin{minipage}[t]{0.23\textwidth}\centering
Karame \\ Mohammadiporshokooh\orcid{0009-0000-8349-3389}$^{1}$\\\small kmoham6@lsu.edu
\end{minipage}
\hfill
\begin{minipage}[t]{0.23\textwidth}\centering
Panagiotis Syskakis\orcid{0009-0005-0594-1445}$^{1}$\\\small psyskakis@lsu.edu
\end{minipage}
\hfill
\begin{minipage}[t]{0.23\textwidth}\centering
Andrew Lumsdaine\orcid{0000-0002-9153-6622}$^{2}$\\\small al75@uw.edu
\end{minipage}
\hfill
\begin{minipage}[t]{0.23\textwidth}\centering
Hartmut Kaiser\orcid{0000-0002-8712-2806}$^{1}$\\\small hkaiser@cct.lsu.edu
\end{minipage}
\\[1em]
$^{1}$Department of Computer Science, Center for Computation \& Technology, \\
 Louisiana State University, Baton Rouge, USA \\
$^{2}$School of Computer Science and Engineering, University of Washington, Seattle, USA
}


\maketitle

\begin{abstract}

Graph processing at scale presents many challenges, including the irregular structure of graphs, the latency-bound nature of graph algorithms, and the overhead associated with distributed execution. While existing frameworks such as Spark GraphX and the Parallel Boost Graph Library (PBGL) have introduced abstractions for distributed graph processing, they continue to struggle with inherent issues like load imbalance and synchronization overhead.
In this work, we present a distributed library prototype and a distributed implementation of three key graph algorithms-Breadth-First Search (BFS), PageRank, and Triangle Counting, using C++ mechanisms from the NWgraph library and leveraging HPX's distributed containers and asynchronous constructs.
These algorithms span the categories of traversal, centrality, and pattern matching, and are selected to represent diverse computational characteristics. 
We evaluate our HPX-based implementations against GraphX and PBGL, showing that a high-performance runtime such as HPX enables the construction of algorithms that significantly outperform conventional frameworks by exploiting asynchronous execution, latency hiding, and fine-grained parallelism in shared memory.
All algorithms in our prototype follow a unified execution model in which local and remote computations are expressed using the same programming abstractions, with asynchrony managed transparently by the runtime. This design explicitly leverages shared-memory parallelism within each locality while overlapping communication and computation across localities, providing a practical foundation for extending this approach to a broader class of distributed graph algorithms.

\end{abstract}



\begin{IEEEkeywords}
Graph algorithms, Distributed, High Performance, Graph Computing
\end{IEEEkeywords}



\section{Introduction}
\label{sec:Intro}

Graphs are powerful abstractions for modeling relationships across various domains, including scientific computing, data analysis, and machine learning/artificial intelligence. As these applications continue to grow in size and complexity, they often surpass the computational and memory capacities of single compute nodes, necessitating a transition to high-performance distributed-memory parallel computing solutions. However, conventional high-performance computing approaches, particularly those effective for dense or sparse linear algebra problems, do not readily adapt to graph computations. 
The irregular structures inherent in graphs and the data-driven nature of graph algorithms result in fine-grained and unpredictable data access patterns, leading to significant runtime overhead due to latency. Moreover, many graph algorithms have inherent sequential dependencies and strict operation orderings that limit parallel execution opportunities and introduce synchronization challenges.
%
%
The rapid growth of graph data in domains such as social networks, biological analysis, and recommendation systems has driven the adoption of distributed graph processing frameworks. However, existing systems continue to face fundamental challenges related to communication overhead, load imbalance, and inefficient resource utilization, particularly for large and irregular graphs~\cite{doi:10.1142/S0129626407002843}.

The NWgraph library has emerged as a flexible and efficient solution for graph computation in C++. It provides a lightweight interface for representing graphs and executing algorithms across various hardware platforms. By integrating NWgraph with HPX, a distributed asynchronous runtime system, this work presents a new distributed graph-processing library prototype addressing the limitations of existing frameworks. HPX's fine-grained parallelism, adaptive resource management, and asynchronous execution model make it well-suited for large-scale graph processing.
Additionally, this paper presents the implementation and evaluation of key graph algorithms, including Breadth-First Search (BFS), PageRank, and Triangle Counting, using a distributed implementation of NWgraph on HPX. Through this integration, we aim to demonstrate the advantages of HPX in minimizing communication overhead, achieving load balancing, and improving resource utilization. Our experiments evaluate the performance of these distributed HPX-based implementations against established distributed frameworks, providing insights into their relative efficiencies.


\section{Background}
\label{sec:Background}


\subsection{Distributed Graph Processing}
\begin{figure}[ht]
    \begin{center}
        \includegraphics[width=0.6\linewidth]{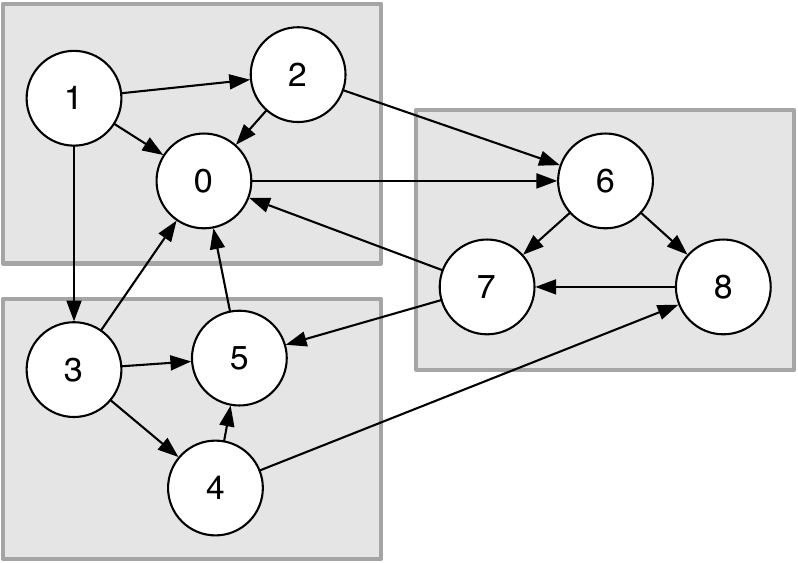}
    \end{center}
\caption{ Structure of a distributed graph.
Each gray box represents a separate locality with its own address space.
As is common in many distributed graph-based applications, 
edges connect from most localities to most others.
}

\label{fig:distributed_graph}
\end{figure}

The nature of distributed graphs is characterized by fine-grained data and control dependencies that span multiple address spaces, tightly coupling computation and communication.
Figure~\ref{fig:distributed_graph} illustrates some of the general difficulties with distributed graph computing by showing a graph partitioned across different 
address spaces (``localities''). Edges in the graph reflect 
fine-grained control and/or data dependencies. In large graphs, many 
edges cross locality boundaries.  Specifically, for a large
random graph, the number of inter-locality edges is
$O(d |V|)$, where $d$ is the average vertex degree.

Algorithms that rely directly on these fine-grained dependencies will incur the latency costs associated with each inter-locality edge.  
Although latency-hiding techniques can mitigate some of these costs, the combination of fine-grained computation and communication, along with the relatively low computation cost compared to communication, remain a fundamental challenge.
Moreover, architectural features that are beneficial in scientific computing applications have little benefit for distributed graph computing.  Many graph problems have little locality or data reuse, rendering caches and prefetching ineffective.  Messages are often sent from every process to almost every other process, negatively impacting communication performance on networks that rely on communications being mostly local (to some small number of neighbors). 



\subsection{Challenges in Graph Processing}

The intrinsic characteristics of graph problems present substantial challenges to effective distributed-memory parallel computing. Graph algorithms are typically developed to optimally address specific problems under theoretically ideal (and generally sequential) conditions.  The subsequent prescription of strict data dependencies and ordering of operations introduces synchronization requirements that can severely limit opportunities for parallelization.  Moreover, execution of graph algorithms is tightly coupled to the structure of the graph, which is only known at runtime.  Graphs are typically irregular, and the data-driven nature of graph algorithms entails fine-grained data accesses to arbitrary locations, incurring significant runtime overhead due to data-access latency.  Because computation in graph algorithms is typically quite small, the ability to hide latency by overlapping communication with computation is limited.

The following are key challenges in distributed graph processing: 

\begin{itemize}
\item \textbf{Communication Overhead:} Distributed graph processing systems experience significant communication overhead as data must be exchanged between nodes for synchronization and state updates~\cite{meng2024surveydistributedgraphalgorithms}. 
The irregular structure of graphs further exacerbates this issue. For instance, algorithms like PageRank require iterative communication, making latency a primary bottleneck ~\cite{doi:10.1142/S0129626407002843}. 
\item \textbf{Load Balancing:} Effective load balancing remains a challenge due to the unpredictable nature of graph workloads even with partitioning techniques. While partitioning techniques like vertex and edge partitioning aim to distribute workloads evenly, they often fail in practice, especially for graphs with skewed degree distributions (e.g., scale-free networks) ~\cite{DBLP:journals/corr/abs-1902-10130}. Poor load balancing results in stragglers, causing overall performance degradation.
\item \textbf{Synchronization Costs:} Many algorithms, such as BFS and Triangle Counting, often involve frequent synchronization points. These barriers cause delays, particularly in large-scale distributed systems ~\cite{Kalavri_2018}. HPX addresses this issue using its fine-grained task parallelism and asynchronous execution model.
\item \textbf{Memory Management:} Managing memory efficiently across distributed nodes is crucial. Graph algorithms often require large memory footprints due to the storage of adjacency lists and intermediate results ~\cite{Sahu_2019}. Conventional frameworks like Pregel and GraphX rely heavily on message passing, which can further strain memory resources. To avoid memory exhaustion, systems are forced to spill data to disk, incurring significant performance penalties due to increased I/O overhead.


\item \textbf{Fault Tolerance:} Distributed systems are prone to node failures, which can severely impact graph processing~\cite{10.1007/s10619-019-07276-9}. Fault tolerance methods include I/O-heavy checkpointing, as employed by Pregel, or lineage-based recovery mechanisms that record how results are derived, as used in systems such as Spark/GraphX, both of which constrain execution to a sequence of super-steps.

\end{itemize}

\subsection{Distributed Graph Tasks}

Graph analytics encompasses a broad range of algorithms for different computational challenges, particularly in large-scale and distributed settings. In this paper, we focus on three fundamental categories, as presented in~\cite{meng2024surveydistributedgraphalgorithms}: 


\textbf{Traversal} algorithms are central to graph analytics, enabling systematic exploration of vertex relationships and substructures. 
Among them, Breadth-First Search (BFS) is widely used due to its simplicity and effectiveness in unweighted graphs. BFS serves as a building block for numerous higher-order algorithms, including shortest path computation and graph connectivity analysis.
It operates in levels, activating a subset of vertices in each iteration (the current frontier), which leads to irregular parallelism and significant load imbalance when frontier sizes fluctuate or vertex degrees vary widely ~\cite{meng2024surveydistributedgraphalgorithms}. Additionally, communication overhead is common in distributed settings due to redundant propagation of visited vertices and synchronization costs between partitions ~\cite{doi:10.1142/S0129626407002843}.



\textbf{Centrality} algorithms are designed to quantify the importance of nodes in a graph based on their position and role in the network, a critical aspect in applications such as web search ranking, social network analysis, and recommendation systems. 
While some centrality measures, such as betweenness and closeness, require global graph traversal and shortest-path calculations, PageRank operates primarily through localized neighbor information, making it more amenable to parallel and distributed environments. Still, the challenges outlined in section \ref{sec:Intro} manifest strongly in PageRank when handling high-degree vertices, which often dominate compute and message volume. Thus, various optimizations have been proposed to improve its scalability on high-performance computing architectures~\cite{turn0search19}.


 \textbf{Subgraph Analysis} focuses on identifying and analyzing specific structural patterns within a graph, which is crucial for detecting communities, understanding network motifs, and evaluating graph connectivity. A fundamental problem in this category is Triangle Counting, which determines the number of closed three-vertex cycles in a graph.
 Due to the irregular nature of real-world graphs, efficiently parallelizing triangle counting in distributed settings presents challenges in balancing workloads and reducing redundant computations~\cite{7284394}. Various techniques, such as edge-based partitioning and graph sparsification, have been introduced to accelerate triangle counting while maintaining accuracy in large-scale graphs. 

 Together, these three algorithmic categories capture a broad spectrum of communication patterns, synchronization requirements, and workload irregularities commonly encountered in distributed graph processing. 
 In the following sections, we use these as representative workloads to evaluate how the combination of NWGraph’s data structures and HPX’s asynchronous execution model enables uniform expression, effective load balancing, and efficient exploitation of local parallelism across diverse distributed graph tasks.

 \subsection{Related Work}

Large-scale graph analytics has motivated the development of a variety of graph processing frameworks. Graph computation libraries can vary in several key aspects, such as the programming model they expose, their flexibility and generality, as well as the runtime and execution characteristics (CPU, GPU, distributed memory).
Table~\ref{table:related_work} gives an overview of features included in various graph library solutions. 
%
\input{related_work_table}

Libraries in distributed memory are built around different communication/synchronization models. For instance, \textbf{Pregel} restricts data exchange to successive computational steps in a model known as Bulk Synchronous Parallel (BSP). Built on Spark’s resilient distributed dataset (RDD) abstraction, \textbf{GraphX} follows a similar BSP interface for graph analytics.
BSP implementations streamline algorithm development and fault-tolerance, but global synchronization can amplify load imbalance.
The BSP model is also prone to memory exhaustion, as localities receive messages during a superstep but typically do not process them or reclaim associated buffers until the end of an epoch.
Other frameworks utilize fine-grained parallelism and active messages to enable fully asynchronous algorithms, interleaving communication and computation.
Both \textbf{Parallel Boost Graph Library (PBGL)} and its enhanced version \textbf{PBGL 2.0} support fundamental graph algorithms, with the latter using active messages to improve scalability. 

Another differentiating factor between graph frameworks is the algorithmic abstraction that they provide. For example, some enforce a vertex-centric approach while others offer a block‑ or linear‑algebra‑based abstraction. \textbf{Vertex-centric}~\cite{vertex_centric} frameworks are designed to overcome the limitations of MapReduce~\cite {MapReduce} in iterative, interdependent tasks by focusing computations at each vertex and enabling direct communication with neighboring vertices. 
Frameworks like \textbf{Pregel} or \textbf{PowerGraph} express algorithms in customizable but predefined steps. Specifically in PowerGraph, all algorithms are implemented in terms of four vertex operations: "gather", "sum", "apply", and "scatter".
In contrast, the proposed \textbf{Distributed GraphBLAS API}~\cite{9150368} expands on the shared-memory GraphBLAS, treating graphs as sparse matrices and expressing computation using linear-algebra primitives.


\section{Technical Approach}

\subsection {NWGraph}
\label{sec:nwgraph}

NWGraph~\cite{nwgraph_github} is a generic library of graph algorithms in C++, demonstrating how the existing capabilities of the language and its standard library can effectively support graph algorithms and data structures. 
It realizes generic algorithms as function templates, and the type requirements as C++ concepts. 
The process centers on defining type requirements at the interfaces to algorithms based on the specific requirements of each algorithm. 
By utilizing C++ concepts, NWGraph systematically lifts unnecessary constraints, and organizes the minimal type requirements into a coherent taxonomy. 
These concepts define the interfaces for the library's algorithms, enabling seamless composition with other independently-developed components.
Another foundational design decision in NWGraph is that the abstract interface presented by graphs is that of a \textbf{range of ranges}: The outer range is a range over the vertices, and the inner ranges are ranges over each vertex’s neighbor edges.

\noindent\textbf{Concept taxonomy:} Using C++ concepts to define adjacency list and edge list, NWGraph allows for the extension or creation of new graph algorithms without being tied to specific underlying data representations. By defining a graph as a random access range of forward ranges, data structure requirements are generalized across multiple algorithms. This generalization facilitates the design of APIs that offer flexibility with memory and data structure types, amenable to scalability across heterogeneous systems. Developers can integrate custom data structures by adhering to specified concept requirements, enabling integration into broader ecosystems and supporting cross-compatibility between different graph processing tools. 

\noindent\textbf{Reusable data access patterns:} NWGraph's range-based approach allows the implementation of a rich set of range adaptors, allowing for converting one representation of a graph into another. This approach eliminates the need for visitor objects, making the adaptation to specific algorithms seamless and more efficient. For example, providing traversal-based views such as BFS and DFS ranges streamlines traversal operations across different graph structures.

\noindent\textbf{Parallel execution:} NWGraph leverages modern C++ execution policies to fully parallelize algorithms, originally utilizing Intel's Threading Building Blocks (TBB), and later HPX, to enhance thread management and optimize workload partitioning. This approach eliminates the need for manually applied locking mechanisms and coarse-grained mutexes, and enables thread-safe data structures and finer control over concurrency granularity.  

%
%

The solid theoretical foundation of NWGraph's building blocks inherently allows for capturing the full range of necessary graph operations, especially in highly irregular and data-driven computations.

\subsection{HPX Runtime System}

\label{sec:hpx}

The HPX runtime system~\cite{hpx_joss_paper,hpx_stellar_group,kaiser_2024_598202} has several features that are key to addressing the design goals listed in Section~\ref{sec:Background}: asynchronous execution, communication and message batching, dynamic partitioning and load-balancing, and portability in code and performance. The primary design objective of HPX is to create a state-of-the-art asynchronous, parallel, many-task runtime system (AMT) that serves as a solid foundation for scalable applications, while remaining as efficient, as portable, and as modular as possible~\cite{10.1007/978-3-031-31209-0_1}. 
HPX provides the infrastructure for: 

\noindent\textbf{Active global address space (AGAS):} HPX implements a single global address space across all localities an application runs on. AGAS allows all global objects to be moved transparently to different localities without changing their global aaddress, facilitating the need for dynamic load-balancing of the distributed graph data structures.
%

\noindent\textbf{Threads and thread management:} The HPX thread manager implements a work queue-based execution model that supports work-stealing. Tasks that are ready to be executed are wrapped in HPX-threads and scheduled to run. HPX-threads are lightweight and cooperatively (non-preemptively) scheduled in user mode by the thread manager, which significantly reduces context switching and scheduling overheads.

\noindent\textbf{Message transport and message management:} Any inter-locality messaging in HPX is based on separate network messages (called Parcels). Those are an extended form of active messages~\cite{10.1145/139669.140382} and enable asynchronous message-driven, distributed control flow, and dynamic resource management. Parcels are either used to move the work to the data or to gather small pieces of data back to the caller.
%

\noindent\textbf{Local and global synchronization objects:} These are abstractions of different functionalities for event-driven creation, organization of asynchronous flow control, protection of data structures from race conditions, automatic event-driven on-the-fly scheduling of work, and synchronization based on the Futures abstraction~\cite{Friedman1976CONSSN}. In addition to specialized synchronization primitives, HPX provides all such abstractions as specified by the C++ standard (mutexes, condition variables, semaphores, etc.) that are usable to cooperatively synchronize with an HPX-thread, without blocking progress of the underlying kernel (p)thread.

\noindent\textbf{Performance Counter Infrastructure:} HPX exposes a uniform and extensible API for extracting performance-related information about the runtime system itself and about the application's behavior. This allows exposing both predefined performance metrics as well as user- or application-defined performance information to be extracted on demand, for making runtime adaptive decisions that change algorithmic and system parameters.



\subsection{Our Approach}

In this paper, we utilize the existing facilities in NWGraph and HPX, extending those to the requirements of distributed graph computation. The fundamental approach of NWGraph is extended into the distributed domain, with the goal of creating an expressive, generic, and performant distributed graph computation framework.



\noindent\textbf{Distributed range-of-ranges:} We create distributed graph data structures that adhere to the NWGraph's range-based approach. 
By substituting the underlying storage to a distributed HPX container (from \texttt{std::vector} to \texttt{hpx::partitioned\_vector}), the abstract interface remains a \textit{range of ranges}. 
This abstraction preserves the uniform algorithm interface, allowing generic graph algorithms written for local NWGraph containers to be instantiated over distributed representations with minimal adaptation. Locality-aware iteration, which is necessary for performant algorithm implementations, is then a refinement of the range interface.

\noindent\textbf{Parallel \& Distributed execution:} For parallelism, we leverage HPX's parallel algorithms and async/futures, all of which closely mirror their C++ standard library counterparts. HPX extends beyond standard C++ with distributed data structures, components, remote actions, distributed futures for asynchronous cross-locality operations, and distributed collective operations, capabilities essential for distributed-memory graph algorithms.

\section{Implementation}

For our initial prototype of a distributed data structure for graphs, we combined the NWGraph data structures with HPX's partitioned vector implementation. This prototype was used to implement distributed versions of Triangle Counting, PageRank, and Breadth-First-Search (BFS) algorithms.
Although these algorithms are generally parallelizable, their performance in distributed systems suffers from high communication overhead, work imbalance, and challenges in data locality. HPX's asynchronous remote procedure call infrastructure was ideal for our needs. 

NWGraph implements (amongst other formats) a Compressed Sparse Row (CSR) representation of graphs.
The CSR format stores vertices and edges in separate arrays, with the indices into these arrays corresponding to the identifier for the vertex or edge, respectively.
The edge array is sorted by the source of each edge, but contains only the targets for the edges.
The vertex array stores offsets into the edge array, providing the offset of the first edge outgoing from each vertex. 
In our prototype, we use \texttt{hpx::partitioned\_vector} as the data store for NWGraph's CSR data structure.
This allows to store the vertices and edges of a graph in a distributed way, such that the vertices and edges are spread across the set of localities (computational nodes) the application is running on.
Then, the edge array is partitioned such that for every vertex, the list of outgoing edges can be traversed locally.

Graph data is represented internally using this partitioned sparse structure, where each partition corresponds to a subset of vertices and edges owned by a locality. While the partition vector provides a global logical view of the graph, the underlying data is distributed across non-contiguous memory regions. This representation naturally aligns with the over-decomposed execution model, enabling independent partitions to be processed concurrently using uniform local computation kernels.

\begin{lstlisting}[
    float,
    caption={Naïve Sequential Triangle Counting Algorithm}, 
    label={lst:naivetc}, 
    escapechar=!
]
template <adjacency_list_graph Graph>
size_t triangle_count(Graph const& G) {
    size_t triangles = 0;
    hpx::for_each(hpx::execution::seq, G.begin(), G.end(), !\label{line:policy}!
        [&](auto u_neighbors) {
            for (auto edge_uv : u_neighbors) {
                auto v = target(G, edge_uv);
                triangles += intersection_size(u_neighbors, G, v);
            }
        });
    return triangles;
}
\end{lstlisting}

HPX's \texttt{hpx::partitioned\_vector} is a distributed data structure representing a segmented array, where the parts of the array are located on different localities of a distributed application.
It exposes an API similar to that of a C++ \texttt{std::vector}, which made the integration with NWGraph straightforward. As such, any C++ algorithm that would work with \texttt{std::vector} readily works with \texttt{hpx::partitioned\_vector}.
e.g., NWGraph's naïve sequential triangle counting algorithm still properly functions when applied to an instance of a distributed graph (see Listing~\ref{lst:naivetc}).
This direct approach, however, shows a significant performance decline in distributed environments, as per-element access potentially requires a synchronous network operation.
%
The distributed algorithm shown in Listing ~\ref{lst:dist_naivetc} mitigates this effect by asynchronously launching the triangle counting on each of the graph's partitions.

HPX implements a full set of the C++ standard (parallel) algorithms that also have special overloads taking advantage of segmented iterators~\cite{10.5555/647373.724070} if invoked on a partitioned vector.
These algorithms asynchronously schedule operations on the separate partitions that are run concurrently on the correct locality.
For our prototype, we have used the HPX segmented algorithm infrastructure for the distributed triangle counting, BFS, and PageRank algorithms, for which we provide performance and scaling results in section \ref{sec:Results}.

All graph algorithms in this work follow a unified execution model in which local and remote computations are expressed using the same programming abstractions. Operations on locally owned graph partitions are executed directly, while operations targeting remote partitions are issued as asynchronous remote actions returning futures. From the algorithm’s perspective, both cases invoke the same local computation routines, with asynchrony handled transparently by the HPX runtime. Synchronization is deferred until global reduction points, ensuring that computation proceeds without blocking while remote operations are in flight.

To mitigate load imbalance and hide communication latency, the graph is decomposed into more partitions than available processing cores on each locality. This intentional over-decomposition allows HPX’s work-stealing scheduler to continue executing ready tasks while other tasks are suspended awaiting remote results. As a result, threads that cannot make progress due to pending communication do not stall the system, and available cores are continuously utilized with independent local work. This design explicitly exploits shared-memory local parallelism within each locality, allowing multiple graph partitions to be processed concurrently using HPX’s lightweight threading and work-stealing scheduler.

In addition, processing resources are explicitly divided between local computation and the execution of incoming remote requests. Reserving a subset of cores for servicing remote actions ensures that distributed operations can make progress promptly, preventing starvation caused by purely local workloads. Although this allocation is static in the current implementation, it provides a practical mechanism for maintaining responsiveness under sustained communication pressure.

Although explicit message batching is not implemented, the use of deferred asynchronous remote actions implicitly enables a limited form of message coalescing. Multiple remote requests issued within a short time window may be combined by the runtime before transmission, reducing communication overhead without requiring algorithm-level buffering or synchronization.

%

\begin{lstlisting}[
    float,
    caption={Naïve Distributed Triangle Counting Algorithm},
    label={lst:dist_naivetc},
    escapechar=!
]
template <adjacency_list_graph Graph>
size_t distributed_triangle_count(Graph const& G, size_t first, size_t last) {
  size_t triangles = 0;
  std::vector<hpx::future<size_t>> counts;

  hpx::for_each(hpx::execution::seq,
      G.begin() + first, G.begin() + last,
      [&](auto u_neighbors) {
        for (auto edge_uv : u_neighbors) {
          auto v = target(G, edge_uv);
          if (vertex_is_on_this_locality(G, v))
            triangles += intersection_size(u_neighbors, G, v);
          else
            counts.push_back(hpx::async(  !\label{line:tcasync}!
              intersection_size,
              vertex_locality(G, v),
              u_neighbors, hpx::ref(G), v));
        }
      });

  return std::transform_reduce(
      counts.begin(), counts.end(), triangles,
      [](hpx::future<size_t> f) { return f.get(); }, !\label{line:fget}!
      std::plus<>());
}
\end{lstlisting}

\textbf{Distributed Triangle Counting} relies on a slightly more complicated algorithm~\cite{DBLP:conf/pasc/KanewalaZL18} (see also Listing~\ref{lst:dist_naivetc}). 

In essence, we perform the intersection of neighbors of the current vertex and its neighbor's neighbors locally if the neighbor's data is placed on the same locality as the original vertex.
In this case, all the information needed for the intersection operation is available on the current locality.
Otherwise, we asynchronously perform the intersection operation on the locality of the vertex's neighbor while passing along the list of neighbors of the original vertex.
Now the other locality will have all the information it requires to perform a fully local intersection operation.
Note that the function \texttt{hpx::async} performs an asynchronous remote procedure call to the given locality (see Listing~\ref{lst:dist_naivetc}, Line~\ref{line:tcasync}).
It immediately returns an instance of an \texttt{hpx::future<size\_t>}, an object that represents the number of triangles identified by the remote operation, even if it has not arrived yet.
Only when calling \texttt{f.get()} on that future, the executing thread will possibly suspend until the value has become available (see Listing~\ref{lst:dist_naivetc}, Line~\ref{line:fget}).

The code shown in Listing~\ref{lst:dist_naivetc} is asynchronously scheduled to run concurrently on each of the partitions of the graph (i.e. on the graph's vertex range \texttt{[first, last)}).
Here we show the sequential version of the algorithm (see Listing~\ref{lst:naivetc}, Line~\ref{line:policy}), however using \texttt{hpx::execution::par} as the first argument to \texttt{hpx::for\_each} will parallelize the loop's execution.
It is interesting to note, that the use of NWGraph's data structures and APIs in combination with HPX's partitioned vector and asynchronous execution management APIs allows to write a distributed algorithm that is very similar to the naïve sequential local algorithm (see Listing~\ref{lst:naivetc}), while scaling out well (see Section~\ref{sec:Results}). 
We also would like to note that the shown algorithm is \textbf{generic} and \textbf{uniform} as it works with other internal graph representations (in addition to the used CSR format). 

\begin{lstlisting}[float,
        caption={Distributed PageRank iteration},
        label={lst:dist_naivedpr},
        escapechar=@
       ]
using Reply = std::tuple<vertex_id, double>;
Reply compute_contribution(vertex_id u, vertex_id v
                           DistArray& pr, DistArray& deg)
{
    return {u, pr[v]/deg[v]};
}

// Runs on the owner of [first,last)
double pagerank_iteration(const DistGraph& G,
                        vertex_id first, vertex_id last,
                        DistArray& pr, DistArray& deg,
                        DistArray& accum, double damping)
{
    std::vector<hpx::future<void>> pending;
    for (vertex_id u = first; u != last; ++u) {
        accum[u] = 0.0;
        for (auto edge_uv : G[u]) {
            vertex_it v = target(G, edge_uv);
            if (is_local(v)) {
                accum[u] += pr[v] / deg[v];
            } else {
                // Request contribution from owner of v
                pending.push_back(
                    hpx::async(compute_contribution,  @\label{line:prasync}@
                        vertex_locality(G, v), 
                        u, v, pr, deg)
                    .then([&](auto&& f){
                        auto [u, contribution] = f.get();
                        atomic_add(accum[u], contribution);    
                    });
                );
            }
        }
    }
    hpx::wait_all(pending);
    std::size_t N = G.size();
    double base = (1.0 - damping) / N;
    double delta = 0.0;
    for (vertex_id u = first; u != last; ++u) {
        double old = pr[u];
        pr[u] = base + damping * accum[u];
        delta += abs(pr[u] - old);
    }
    return delta;
}
\end{lstlisting}

\textbf{Distributed PageRank} is implemented following the formulation of the random surfer model proposed by Brin and Page~\cite{brin1998anatomy}, where the PageRank score for vertex $u$ is computed as:

\[
PR_{\text{new}}(u)=\frac{1-d}{N}+d\cdot A(u)
\]
 where $d$ is the damping factor (typically set to 0.85), $N$ is the total number of nodes, and
\[
A(u)=\sum_{v\in N_{\text{in}}(u)} \frac{PR(v)}{\deg(v)}.
\]
where $N_{\text{in}}(u)$ represents the set of in-neighbors of $u$, and $deg(v)$ denotes the out-degree of vertex $v$.

%

A simplified version of our implementation is shown in Listing \ref{lst:dist_naivedpr}. Along with the distributed adjacency list, the algorithm maintains: i) a distributed PageRank vector $PR$ (one value per vertex), ii) a distributed out-degree vector $\deg$ (one value per vertex), and iii) a distributed accumulator vector $A$ used to collect incoming contributions during an iteration. An iteration is launched on each compute node that owns a graph partition, and performs one PageRank relaxation step; For each locally-owned vertex $u$, the worker traverses $u$’s adjacency list and accumulates neighbor contributions into $A(u)$. If a neighbor vertex $v$ is locally owned, the contribution $\frac{PR(v)}{\deg(v)}$ is computed directly from local state and added to $A(u)$. Otherwise, an asynchronous remote action is launched on the owner of $v$, which computes and provides the contribution derived from remote vertex $v$ (see Listing~\ref{lst:dist_naivedpr}, Line~\ref{line:prasync}). This way, computation is scheduled so that the contribution is computed on the node that owns the data (“move compute to data”), minimizing remote memory traffic. Finally, we want to note the use of \texttt{.then}, used to attach a continuation to the asynchronous remote operation. This way, the local data is updated as soon as the remote contribution arrives from the remote node.





\textbf{Distributed BFS} is implemented similarly 
as a traversal that builds a distributed parent tree rooted at a user-provided source vertex. 
A complete implementation of this distributed BFS algorithm, including code and experimental evaluation on different graph datasets, was previously presented in our earlier work~\cite{mohammadiporshokooh2026initialevaluationdistributedgraph}.

\section{Experimental Measurements and Results}
\label{sec:Results}

\begin{figure*}[ht]
    \centering
    \subfloat[Distributed Triangle Count\label{fig:dtc_size}]{
        \includegraphics[width=0.80\textwidth]{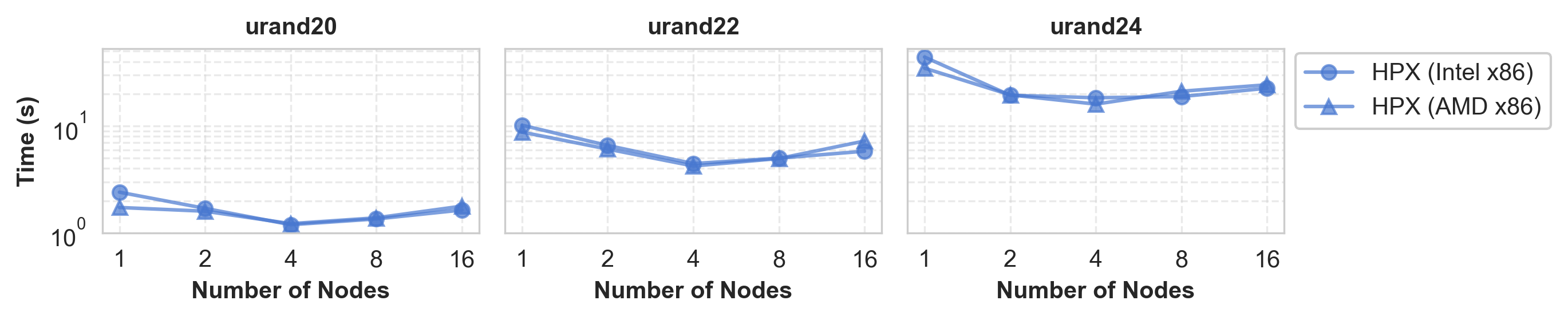}
    }\\
    \subfloat[Distributed BFS\label{fig:dbfs_size}]{
        \includegraphics[width=0.80\textwidth]{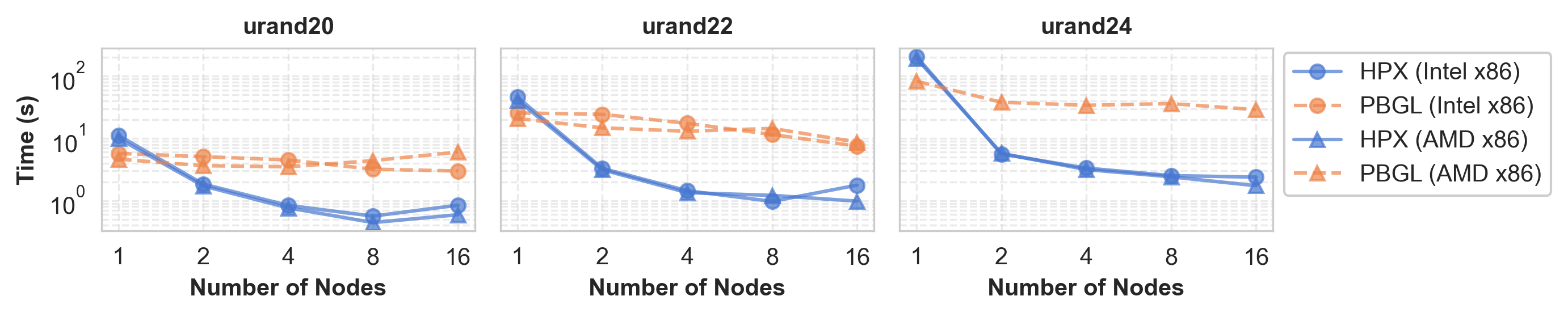}
    }\\
    \subfloat[Distributed PageRank\label{fig:dpr_size}]{
        \includegraphics[width=0.80\textwidth]{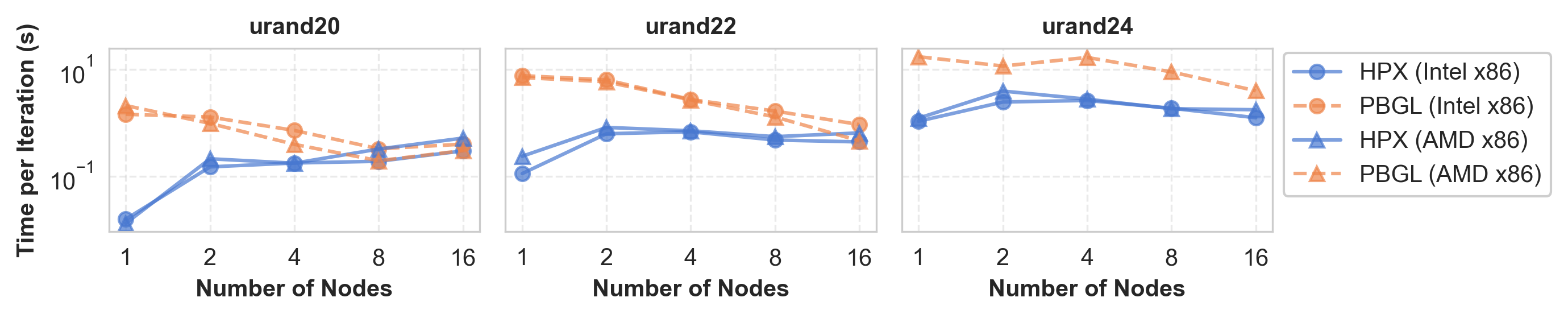}
    }
    \caption{Strong scaling behavior of distributed graph algorithms on Erd\"os-R\'enyi uniformly random (\texttt{urand}) graphs.}
    \label{fig:urand_by_size}
\end{figure*}


For measuring the performance of our implementations, we tested our algorithms on sample graphs from the GAP-dataset \cite{beamer2017gapbenchmarksuite}, as well as synthetically generated graphs using the Erd\"os-R\'enyi "urand" model. The \texttt{urand} graphs are of varying size, with an average vertex degree of 32 (e.g., urand20 has $2^{20}$ vertices and $2^{25}$ edges). The programs were compiled using GCC 14.2.0, and measurements were taken on two different clusters: The first cluster consists of 16 nodes, each utilizing two 20-core Intel(R) Xeon(R) Gold 6148 CPU processors and 92 GB of memory. The second cluster also consists of 16 nodes, each utilizing two 24-core AMD EPYC 7352 24-Core processors and 264 GB of memory.

Figure \ref{fig:urand_by_size} shows the execution time of the three algorithms as the number of computation nodes increases. We compare the performance of our BFS and PageRank implementations to the reference distributed implementations provided by the Parallel Boost Graph Library (PBGL) (a reference Triangle Count implementation is not provided by PBGL). Distributed BFS shows a good scaling behavior on the HPX implementation (blue line), surpassing the performance of the PBGL implementation by an order of magnitude on 8 and 16 nodes (Fig.  \ref{fig:dbfs_size}). Notice that going from 1 to 4 nodes, the HPX implementation exhibits super-scaling behavior, as the parts of the distributed graph start fitting in the CPU caches. Distributed PageRank does not generally scale as well, with the HPX implementation surpassing the performance of PBGL on fewer machines (\ref{fig:dpr_size}). The performance of the two implementations seems to converge when running on many nodes, suggesting that the execution time is dominated by communication latencies and synchronization overheads. We also show the results for the HPX distributed triangle count implementation, where our highly asynchronous approach seems to scale up to 4 nodes on both machines (Fig. \ref{fig:dtc_size}).

As observed, there are several factors affecting the scaling behavior of our implementation. Firstly, as a graph is distributed to more localities, the ratio of remote to local vertices increases. This leads to more computations involving a remote operation, instead of being resolved locally. For smaller graphs, the effectiveness of local parallelism may also decrease, as the workload per thread becomes too small to amortize the parallelization overhead. In contrast, the distribution of data may lead to positive effects when it comes to memory and cache behavior - each computation node is responsible for a small subset of the graph, making cache hits more probable.

\begin{figure}[ht]
    \centering
    \includegraphics[width=1.0\linewidth]{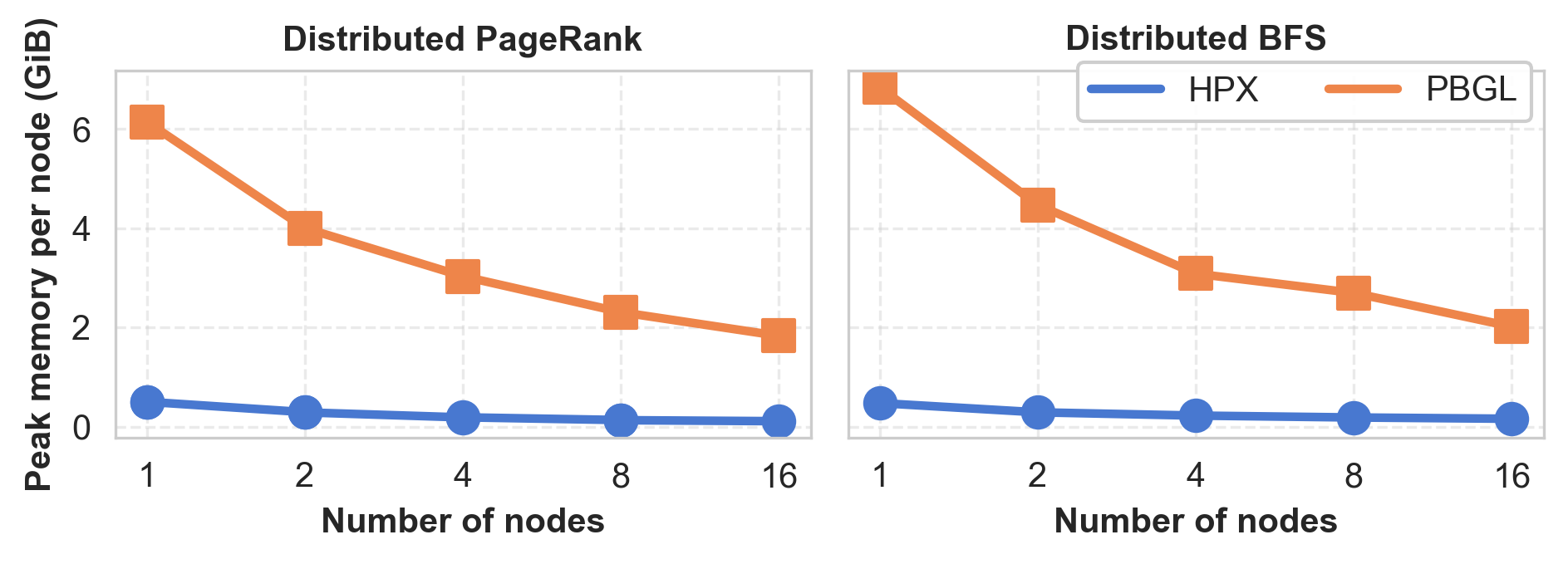}
    \caption{Per-node memory usage comparison using 4 MPI processes (PBGL) / 4 worker threads (HPX) per node.}
    \label{fig:memory_comparison}
\end{figure}

In our experiments, we often faced PBGL suffering from memory exhaustion, explaining the missing data for larger graphs in Figure \ref{fig:urand_by_size}. Figure \ref{fig:memory_comparison} compares the per-node memory usage of the PBGL and HPX implementations, using 4 cores per node. In PBGL, local parallelism is achieved using multiple processes on separate address spaces, which communicate using message-passing (MPI). This causes large data duplication that is proportional to the amount of parallelism. The over-eager caching of remote ("ghost") graph data also increases the memory footprint. In contrast, HPX relies on shared-memory parallelism within a node, so its memory consumption remains nearly constant as the number of cores increases.

\begin{figure*}[ht]
    \centering
     \subfloat[Distributed Pagerank \label{fig:graphx_dpr}]{
    \includegraphics[width=0.9\linewidth]{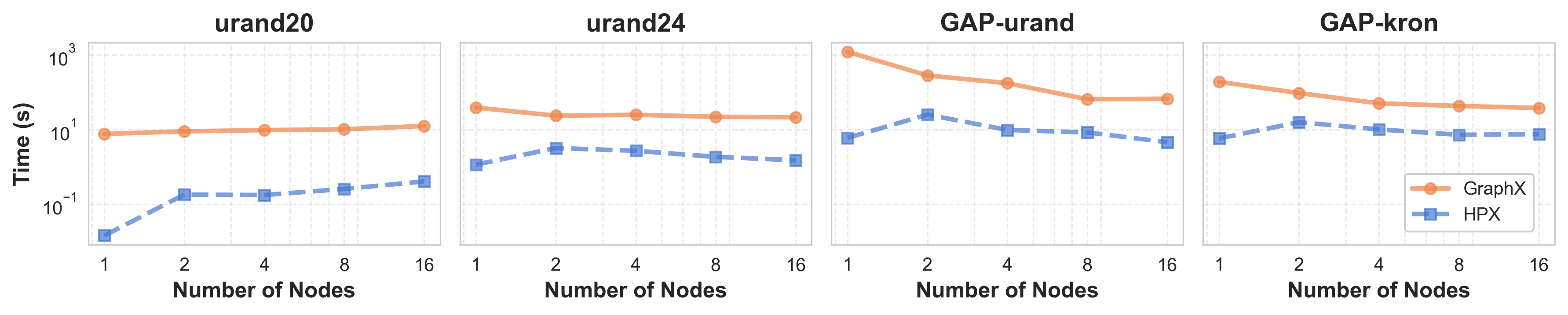}
    }\\
    \subfloat[Distributed Triangle Count \label{fig:graphx_dtc}]{
    \includegraphics[width=0.9\linewidth]{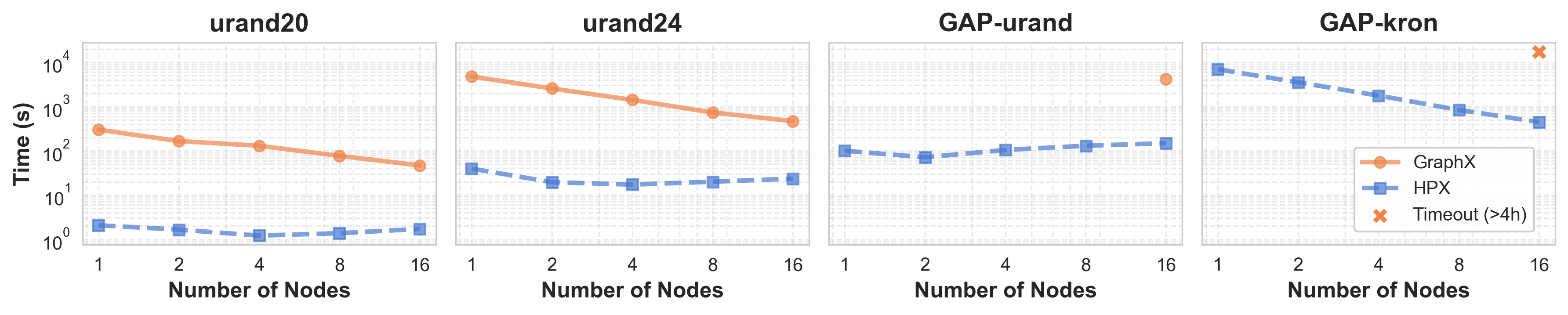}
    }
    \caption{Strong scaling against Spark GraphX implementations, including GAP-kron and GAP-urand graphs ($\sim$128M vertices and $\sim$4B edges).}
    \label{fig:results_graphx}
\end{figure*}

In Figure \ref{fig:results_graphx}, we also compare our implementations of the distributed BFS and distributed PageRank algorithms against the Spark GraphX library, allowing our experiments to extend to larger graphs from the GAP dataset. GAP-urand and GAP-kron are synthetically generated graphs consisting of $2^{27}$ ($\sim$128 M) vertices and $2^{32}$ ($\sim$4B) edges. In contrast to the Erd\"os-R\'enyi "urand" model, the Kronecker graphs produce heavy‑tailed degree distributions closer to what is seen in many real networks (few high‑degree, many low‑degree nodes). GraphX also relies on separate processes for parallelism, and is more I/O-intensive due to Spark's RDD-based fault-tolerant computation model. Our in-memory asynchronous implementation is faster, often by orders of magnitude. On larger graphs, GraphX often exceeded our allowed cluster time limits, and intermediate algorithm data often depleted our memory or even disk storage limits.

The observed performance improvements arise from three design choices enabled by HPX: the use of asynchronous remote function invocations to overlap communication and computation, the elimination of global synchronization barriers, and fine-grained shared-memory parallelism with work stealing to efficiently utilize CPU cores and memory bandwidth. These results highlight the appeal of HPX as a C++ runtime for high-performance distributed graph computation.



The results in Figures~\ref{fig:urand_by_size} and~\ref{fig:results_graphx} come from prototypes that have not exhausted the space of possible optimizations. For instance, messages can be coalesced either in code or by the runtime; 
partition sizes could be adaptively chosen for better load balancing; and shuffling partition and vertex distributions across nodes may enhance load balancing. For local work, blocking techniques and cache-aware traversal can greatly improve cache utilization \cite{Zhang2017Caches8257937}. For simplicity, any specialized graph partitioning (e.g., minimizing cross-partition edges) was also omitted.


\section{Future Work}


The prototype implementations and experimental results presented in this work expose several promising directions for future research. In particular, while our current design already benefits from asynchronous execution, locality-aware computation, and limited forms of message aggregation, a systematic exploration of runtime adaptivity remains an important next step. Many performance-critical parameters, such as message coalescing thresholds, partition sizes, and vertex-to-locality placement, are currently fixed or heuristically chosen. Dynamically adapting these parameters at runtime raises several open research questions, including: How many messages should be accumulated before issuing a coalesced communication? What partition sizes best balance communication overhead and load flexibility? How should vertices and partitions be distributed to evenly balance asynchronous remote operations? Additionally, it remains an open question whether parallelizing certain outer loops is always beneficial, or whether sequential execution may be preferable when computation is already distributed across partitions.

Additionally, our prior work has demonstrated the effectiveness of runtime adaptivity for processing-unit selection and chunk-size control in HPX’s parallel algorithms~\cite{mohammadiporshokooh2025new, 10.1007/978-3-031-97196-9_6}. In that work, adaptive executors dynamically tune the number of active cores and the granularity of work distribution based on runtime measurements, yielding consistent performance improvements across diverse workloads. A natural and promising extension of the present work is to integrate these adaptive mechanisms into our distributed graph implementations. 


Another important direction is the integration of GPU acceleration into our distributed graph framework. While this requires extending HPX’s execution and data-management infrastructure to better support heterogeneous devices, we expect GPU offloading to further improve throughput for compute-intensive graph kernels.


Finally, we expect that more complex distributed graph algorithms can be composed in a manner closely resembling their textbook formulations, lowering the barrier for non-experts. This observation motivates future exploration of specialized LLM-based tools that assist in generating and customizing distributed graph code, an idea supported by the simplicity of our BFS and PageRank implementations.

\section{Conclusion}

In this paper, we presented a distributed graph analytics framework built on top of the HPX runtime system and NWGraph data structures. We focused on three fundamental classes of graph algorithms-traversal, centrality, and subgraph analysis-and implemented distributed versions of Breadth-First Search, PageRank, and Triangle Counting using a uniform programming model that closely follows their sequential formulations.


A key contribution of this work is showing that distributed graph algorithms can be expressed in a generic and maintainable way in C++, by integrating partitioned data structures with asynchronous remote task execution. This approach enables computation to be moved to the data while preserving locality and avoiding global synchronization barriers.
Overall, this work demonstrates that HPX can provide a flexible and efficient foundation for scalable distributed graph analytics, opening the door to future research on adaptive execution strategies, heterogeneous architectures, and automated generation of distributed graph algorithms.

\input{sample-base.bbl}
\end{document}

%% file: preamble.tex
\usepackage{wrapfig}
\usepackage{soul}
\usepackage{caption}
\usepackage{hyperref}
\usepackage{placeins}

\usepackage{tabularray}

\usepackage[table,xcdraw]{xcolor}
\usepackage{tabularx}
\usepackage{tikz}

%% file: related_work_table.tex

\newcommand{\pcirclesizeouter}{0.8ex}
\newcommand{\pcirclesizeinner}{0.7ex}

\NewDocumentCommand{\circlehalf}{}{%
    \begin{tikzpicture}
    \fill[black] (0,0) circle (\pcirclesizeouter);
    \fill[white] (0,0) -- (180:\pcirclesizeinner) arc (180:0:\pcirclesizeinner) -- cycle;
    \end{tikzpicture}
}
\NewDocumentCommand{\circlefull}{}{%
    \begin{tikzpicture}
    \fill[black] (0,0) circle (\pcirclesizeouter);
    \end{tikzpicture}
}
\NewDocumentCommand{\circleempty}{}{%
    \begin{tikzpicture}
    \fill[black] (0,0) circle (\pcirclesizeouter);
    \fill[white] (0,0) circle (\pcirclesizeinner);
    \end{tikzpicture}
}


\definecolor{Mercury}{rgb}{0.96,0.96,0.96}
\definecolor{Eggshell}{rgb}{0.98,0.98,0.90}
\begin{table}[ht]
\footnotesize
\centering
\begin{tblr}{
  width = \linewidth,
  colspec = {Q[250]Q[200]Q[20]Q[20]Q[20]Q[20]},
  columns = {c},
  rows = {c},
  column{1} = {l},
  row{2} = {b},
  row{3} = {Mercury},
  row{5} = {Mercury},
  row{7} = {Mercury},
  row{9} = {Mercury},
  row{11} = {Mercury},
  row{13} = {Mercury},
  hline{2} = {3}{l},
  hline{2} = {5}{},
  hline{2} = {4}{},
  hline{2} = {6}{r},
}
\hline
                                                &                                           & \SetCell[c=4]{} Features &&   &               &               \\ 
                                                & Type                                      & DM            & AS            & HPR           & GA            \\
\hline                                                                                                                                                                                
PBGL 2.0 \cite{edmonds2022parallel}    & Generic graph library                     & \circlefull   & \circlefull   & \circlehalf   & \circlefull   \\
PBGL \cite{PBGL}                                & Generic graph library                     & \circlefull   & \circleempty  & \circleempty  & \circlehalf   \\
BGL \cite{siek_2001_boost}                      & Generic graph library                     & \circleempty  & \circlefull   & \circleempty  & \circlefull   \\
NWGraph \cite{Lumsdaine2021NWGraphAL}           & Generic graph library                     & \circleempty  & \circlefull   & \circlehalf   & \circlefull   \\
GraphX \cite{10.1145/2484425.2484427}           & Graph processing framework                & \circlefull   & \circleempty  & \circlehalf   & \circlehalf   \\
PowerGraph \cite{10.5555/2387880.2387883}       & Graph processing framework                & \circlefull   & \circlefull   & \circlehalf   & \circlefull   \\
EAGM \cite{abstract-graph-machine}              & Model for graph algorithms                & \circlefull   & \circlehalf   & \circleempty  & \circlefull   \\
Pregel \cite{10.1145/1807167.1807184}           & Graph processing framework                & \circlefull   & \circleempty  & \circleempty  & \circlefull   \\
GraphBLAS \cite{graphblas}                      & Graph sparse linear algebra               & \circleempty  & \circlefull   & \circlefull   & \circlefull   \\
Ligra  \cite{DBLP:conf/ppopp/ShunB13}           & Graph processing framework                & \circleempty  & \circleempty  & \circleempty  & \circlefull   \\
\hline
\end{tblr}
\caption{\footnotesize Overview of features of various graph library solutions (DM: Distributed Memory, AS: Asynchronous, HPR: High-Performance Runtime, GA: Generic Algorithms)} 

\label{table:related_work}
\end{table}